\documentclass[12pt]{article}
\usepackage{graphicx}
\usepackage{lineno}

\newcommand\pubnumber{SNSN-323-63}
\newcommand\pubdate{\today}

\def\napoli{Laboratoire de l'Acc\'el\'eratoire Lin\'eaire\\ 
Universit\'e Paris-Sud 11\\
B\^atiment 200, 91898 Orsay cedex - France}

\def\Title#1{\begin{center} {\Large #1 } \end{center}}
\def\Author#1{\begin{center}{ \sc #1} \end{center}}
\def\Address#1{\begin{center}{ \it #1} \end{center}}

\newcommand\pubblock{\rightline{\begin{tabular}{l} \pubnumber\\
         \pubdate  \end{tabular}}}
\newenvironment{Abstract}{\begin{quotation}  }{\end{quotation}}
\newenvironment{Presented}{\begin{quotation} \begin{center} 
             PRESENTED AT\end{center}\bigskip 
      \begin{center}\begin{large}}{\end{large}\end{center} \end{quotation}}

\def\beq{\begin{equation}}
\def\eeq#1{\label{#1}\end{equation}}
\def\eeqn{\end{equation}}

\def\beqa{\begin{eqnarray}}
\def\eeqa#1{\label{#1}\end{eqnarray}}
\def\eeqan{\end{eqnarray}}

\let\bar=\overbar

\def\Dslash{\not{\hbox{\kern-4pt $D$}}}
\def\dslash{\not{\hbox{\kern-2pt $\del$}}}

\def\msb{{\bar{\ssstyle M \kern -1pt S}}}

\begin{document}


\begin{titlepage}
\pubblock

\vfill
\Title{Rare Decays with Missing Energy at SuperB}
\vfill
\Author{Alejandro P\'erez}
\Address{\napoli}
\vfill
\begin{Abstract}
The study of rare B-decays at SuperB provides unique opportunities to understand the Standard Model (SM) 
and to constrain new physics (NP). It is discussed the new physics potential of the $B\rightarrow K\nu\bar{\nu}$ 
and $B\rightarrow K^{*}\nu\bar{\nu}$ system from the proposed SuperB experiment with $75{\rm ab^{-1}}$ of data 
(5 nominal years of data taking).
\end{Abstract}
\vfill
\begin{Presented}
6th International Workshop on the CKM Unitary Triangle \\
Warwick, England,  September 6--10, 2010
\end{Presented}
\vfill
\end{titlepage}
\def\thefootnote{\fnsymbol{footnote}}
\setcounter{footnote}{0}

\section{Introduction}

Rare decays with a $\nu\bar{\nu}$ pair in the final state are interesting probes of NP, 
since they allow one to transparently study $Z$ and other electroweak penguin effects in the 
absence of dipole operator and Higgs penguin contributions,  which are often more important than $Z$ 
contributions in $b\rightarrow s \ell^+\ell^-$ decays. Moreover, since the neutrinos escape the detector 
unmeasured, the $B\rightarrow K^{(*)} + E_{\rm miss}$ channel can also contain contributions from other 
light SM-singlet particles substituting the neutrinos in the decay.

Even though the inclusive $B\rightarrow X_s\nu\bar{\nu}$ decay is theoretically the cleanest among the 
$b \rightarrow s \nu\bar{\nu}$ transitions, it is experimentally very challenging to measure. For this reason 
this decay mode is excluded from the following discussion. Out of the two $B\rightarrow K\nu\bar{\nu}$ 
and $B\rightarrow K^*\nu\bar{\nu}$ decay modes, there are three observables accessible: the corresponding branching 
fractions and an additional observable related to the angular distribution of the $K^*$ decay products $K\pi$: the $K^*$ 
longitudinal polarization fraction $\langle F_L\rangle$~\cite{Buras}
\begin{equation}
\label{eq:Angular_distribution}
\frac{d\Gamma}{d\cos\theta} \propto \frac{3}{4}(1-\langle F_L\rangle)\sin^2\theta + \frac{3}{2}\langle F_L\rangle \cos^2\theta ~,
\end{equation}where $\theta$ (helicity angle), is the angle between the $K^*$ direction in the $B$ rest frame and the 
$K$ direction in the $K^*$ rest frame. These three observables only depend on two combinations of the Wilson coefficients 
$C^{\nu}_L$ and $C^{\nu}_R$~\cite{Buras}, through the variables $\epsilon = \sqrt{|C^{\nu}_{L}|^2+|C^{\nu}_{R}|^2}/|(C^{\nu}_{L})^{SM}|$ and 
$\eta = -{\rm Re(C^{\nu}_{L}C^{\nu*}_{R})}/(|C^{\nu}_{L}|^2+|C^{\nu}_{R}|^2)$ ($\eta \in [-\frac{1}{2},\frac{1}{2}]$). 
The discussed observables can be expressed in terms of $\epsilon$ and $\eta$ as follows,
\begin{eqnarray}
\label{eq:Observables}
{\rm Br}(B\rightarrow K^*\nu\bar{\nu})  &=& {\rm Br}(B\rightarrow K^*\nu\bar{\nu})_{SM}(1+1.31\eta)\epsilon^2~, \\
{\rm Br}(B\rightarrow K \nu\bar{\nu})   &=& {\rm Br}(B\rightarrow K \nu\bar{\nu})_{SM}(1-2\eta)\epsilon^2~, \\
\langle{\rm F}_L(B\rightarrow K^*\nu\bar{\nu})\rangle &=& \langle{\rm F}_L(B\rightarrow K^*\nu\bar{\nu})\rangle_{SM}\frac{(1+2\eta)}{(1+1.31\eta)}~.
\end{eqnarray} As $\epsilon$ and $\eta$ can be calculated in any model, these four expressions can be considered 
as fundamental formulae for any phenomenological analysis of the decays in question. The experimental bounds on 
the branching ratios (see table~\ref{tab:Expe_measurements}) can be translated to excluded areas in the 
$\epsilon-\eta$ plane, where the SM corresponds to $(\epsilon,\eta) = (1,0)$ (see the green area of rightmost 
plot in figure~\ref{fig:EpsEta_constrain}). Since the three observables only depend on two parameters, a measurement 
of all of them would overconstrain the resulting $(\epsilon,\eta)$ point.

\begin{table}[t]
\begin{center}
\begin{TableSize}
\begin{tabular}{l|cc}
Observable         & SM prediction & Experiment \\
\hline
${\rm BR}(B\rightarrow K\nu\bar{\nu})$                   & $(6.8^{+1.0}_{-1.1})\times 10^{-6}$~\cite{Buras} & $< 80\times 10^{-6}$~\cite{BaBar_meas} \\
${\rm BR}(B\rightarrow K^*\nu\bar{\nu})$                 & $(4.5 \pm 0.7)\times 10^{-6}$~\cite{Buras}       & $< 14\times 10^{-6}$~\cite{Belle_meas} \\
$\langle{\rm F}_L(B\rightarrow K^*\nu\bar{\nu})\rangle$  & $0.54 \pm 0.01$~\cite{Buras}                     & --- \\
\hline
\end{tabular}
\caption{\em SM predictions and experimental $90\%~{\rm C.L.}$ upper bounds for the four $b \rightarrow s \nu\bar{\nu}$ observables.}
\label{tab:Expe_measurements}
\end{TableSize}
\end{center}
\end{table}

\section{The Experimental Technique and Strategy}

The recoil technique has been developed in order to search for rare $B$ decays with undetected particles, 
like neutrinos, in the final state. The technique consists on the reconstruction of one of the two B mesons 
($B_{\rm tag}$), produced through the $e^+e^- \rightarrow \Upsilon(4S) \rightarrow B\bar{B}$ resonance,
 in a high purity hadronic or semi-leptonic final states, allowing to built a pure sample of 
$B\bar{B}$ events. Having identified the $B_{\rm tag}$, everything in the rest of the event (ROE) belongs by 
default to the signal B candidate ($B_{\rm sig}$), and so this technique provides a clean environment to 
search for rare decays.

In this analysis, the $B_{\rm tag}$ is reconstructed in the hadronic modes 
$B\rightarrow D^{(*)}X$, where $X = n\pi + mK + pK^0_S + q\pi^0$ ($n+m+p+q < 6$), or semi-leptonic modes 
$B\rightarrow D^{(*)}\ell\nu$, ($\ell = e,~\mu$). In the search for $B\rightarrow K\nu\bar{\nu}$ decays, the signal 
is given by a single track identified as a kaon in the rest of the event. In the search of 
$B\rightarrow K^*\nu\bar{\nu}$ decays, it is searched for a single $K^*$ in the ROE reconstructed in the 
$K^{*0} \rightarrow K^+\pi^-$, $K^{*+} \rightarrow K^0_S\pi^+$ and $K^{*+} \rightarrow K^+\pi^0$ modes.

For this kind of decay modes with undetected particles in the final state, the most powerful variable 
for separating signal and background is the so-called extra energy, $E_{\rm extra}$, which is defined 
as the extra energy in the electromagnetic calorimeter not associated with the $B_{\rm tag}$ or 
$B_{\rm sig}$ candidates. For the signal this variable peaks strongly near zero. This variable 
can be combined with the helicity angle to perform a 2-dimensional fit to extract the 
$B\rightarrow K^*\nu\bar{\nu}$ rate and polarization fraction.

In order to perform the angular analysis for the $B\rightarrow K^* \nu\bar{\nu}$ decay it is needed 
the reference frame of the $B_{\rm sig}$. Due to the closed kinematic of the hadronic recoil technique, 
the $B_{\rm sig}$ rest frame can be easily calculated from the reconstructed $B_{\rm tag}$ and the beam 
energies. However, the semi-leptonic recoil technique poses a problem due to the presence of a neutrino 
in the $B_{\rm tag}$ reconstruction. As the only missing particle in the $B_{\rm tag}$ is a neutrino, it 
is possible to calculate the CM-frame angle between the $B_{\rm tag}$ and $D^{(*)}\ell$ momenta. Yet, as the 
$B_{\rm sig}$ and $B_{\rm tag}$ are back-to-back in the CM frame, this means that the $B_{\rm sig}$ 
momentum is contained in a cone around the $D^{(*)}\ell$ system. Using this information, the module of 
the $B_{\rm sig}$ CM momentum ($p^*_{B} = \sqrt{(E^*_{\rm beam}/2)^2 - m^2_{B}}$, with $E^*_{\rm beam}$ 
the total beam energy in the CM-frame) and the beams collision point (beam-spot), it is possible to built 
two estimators of the helicity angle: 1) Average method: the helicity angle is the arithmetic average of the 
helicity angles calculated using all possible $B_{\rm sig}$ directions around the $D^{(*)}\ell$ system; 
2) Beam-spot method: one computes the helicity angle using the $B_{\rm sig}$ direction around the $D^{(*)}\ell$ 
system that gives the minimum distance between the beam-spot and the line that passes through 
the $D^{(*)}\ell$ vertex with the $B_{\rm sig}$ direction.

\begin{table}[t]
\begin{center}
\begin{TableSize}
\begin{tabular}{l|ccc}
$K^*$ decay modes                 & Average (semi-lep) & beam-spot (semi-lep) & hadronic \\
\hline
$K^{*0} \rightarrow K^+\pi^-$     & $11.5$             & $13.9$               & $2.3$ \\
$K^{*+} \rightarrow K^0_S\pi^+$   & $25.3$             & $28.1$               & --- \\
$K^{*+} \rightarrow K^+\pi^0$     & $24.1$             & $26.0$               & --- \\
\hline
\end{tabular}
\caption{\em Resolution on the helicity angle estimated from the SuperB fast simulation. The angles are in degrees.}
\label{tab:Resolution_hel_angles}
\end{TableSize}
\end{center}
\end{table}

The SuperB~\cite{SuperB} fast simulation has been used to estimate the resolution effects on the helicity angle, the results are shown in 
table~\ref{tab:Resolution_hel_angles}. For the semi-leptonic technique, it should be noted that the average method gives 
better results than the beam-spot one. The resolution effects depends strongly on the $K^*$ mode, one obtains the best results 
for the $K^{*0}\rightarrow K^+\pi^-$ decay, due to the lower rate of fake $K^*$ reconstruction. As expected, better results are 
obtained with the hadronic recoil technique due to the closed kinematics. Currently, there are no results on the expected 
sensitivities on the $\langle F_L \rangle$ parameter for the SuperB expected statistics, but studies are on going.

\section{SuperB detector and expected sensitivities}

Even though the expected SuperB~\cite{SuperB} increase in the instantaneous luminosity of a factor of $~100$ already promises significant 
improvements on the before mentionned rare decays, additional activities for detector optimization are currently 
ongoing. The baseline SuperB detector configuration is very similar to BaBar but the boost ($\beta\gamma$) 
is reduced from 0.56 to 0.28. This boost reduction increases the geometrical acceptance 
and so the reconstruction efficiency. Additionally, it is considered the inclusion of a highly performant particle 
identification device (Fwd-PID) based on time-of-flight measurements in the forward region ($17-25$ degrees 
in polar angle).

\begin{figure}[htb]
\centering
\includegraphics[height=3.3cm]{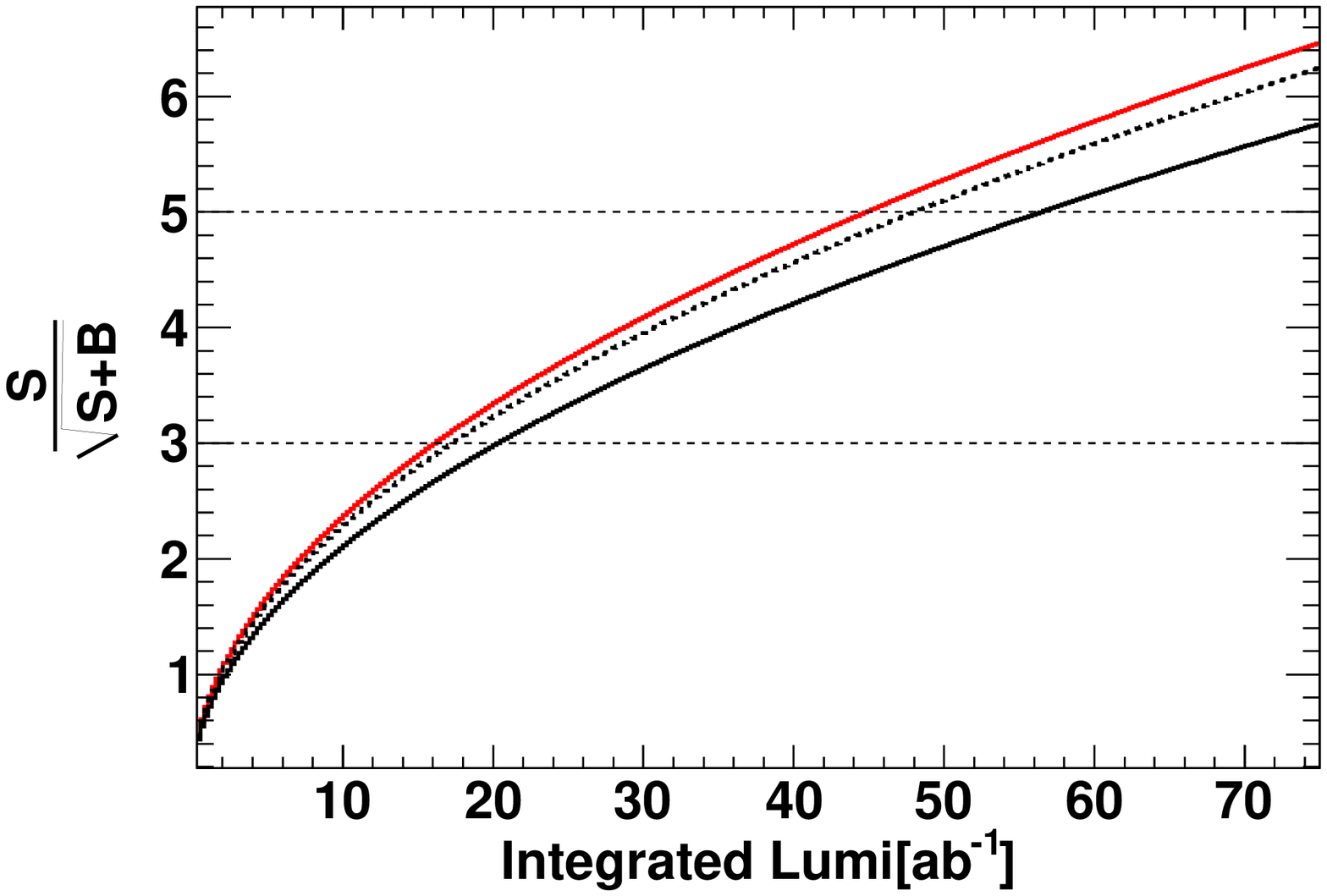}
\includegraphics[height=3.3cm]{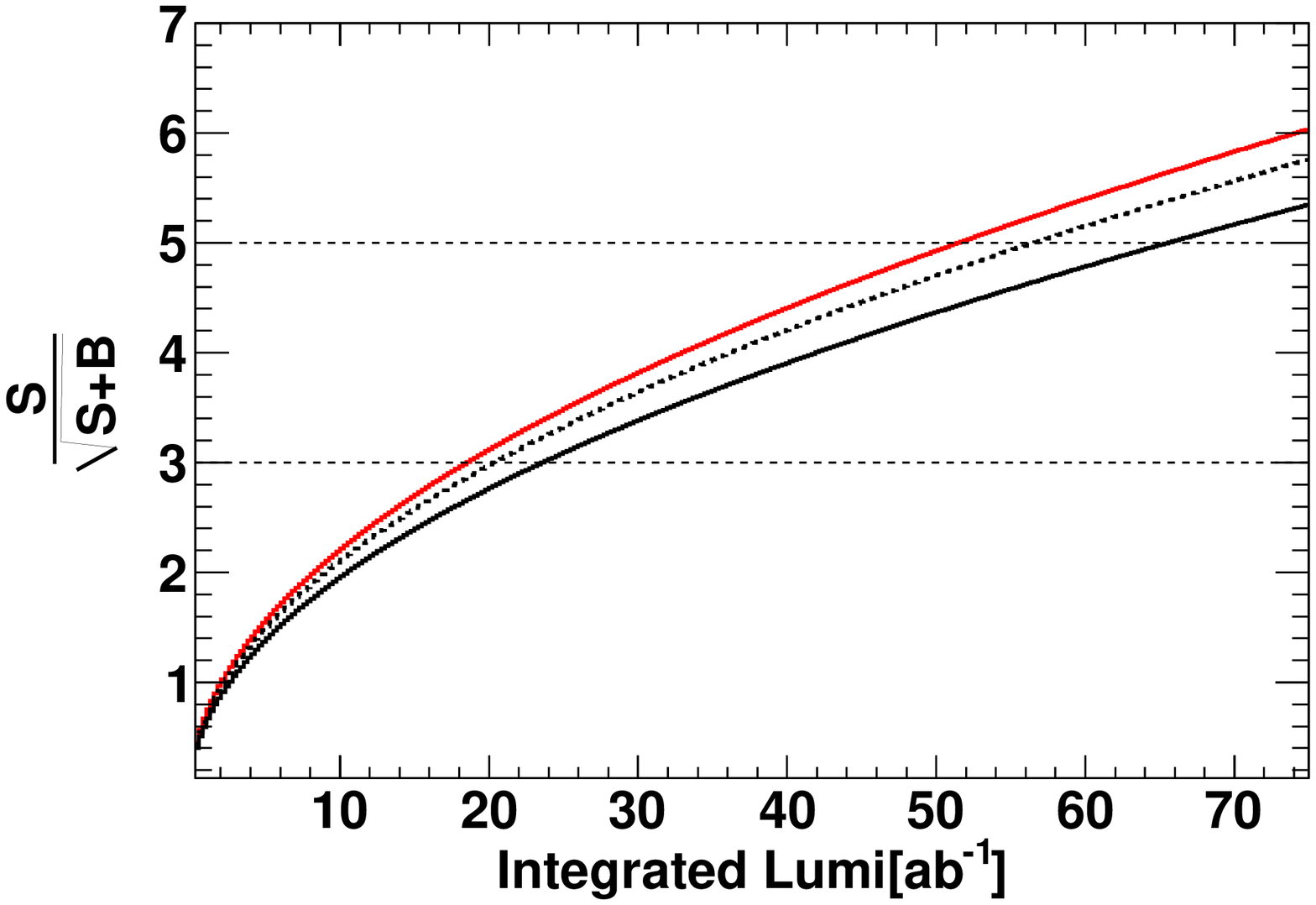}
\includegraphics[height=3.3cm]{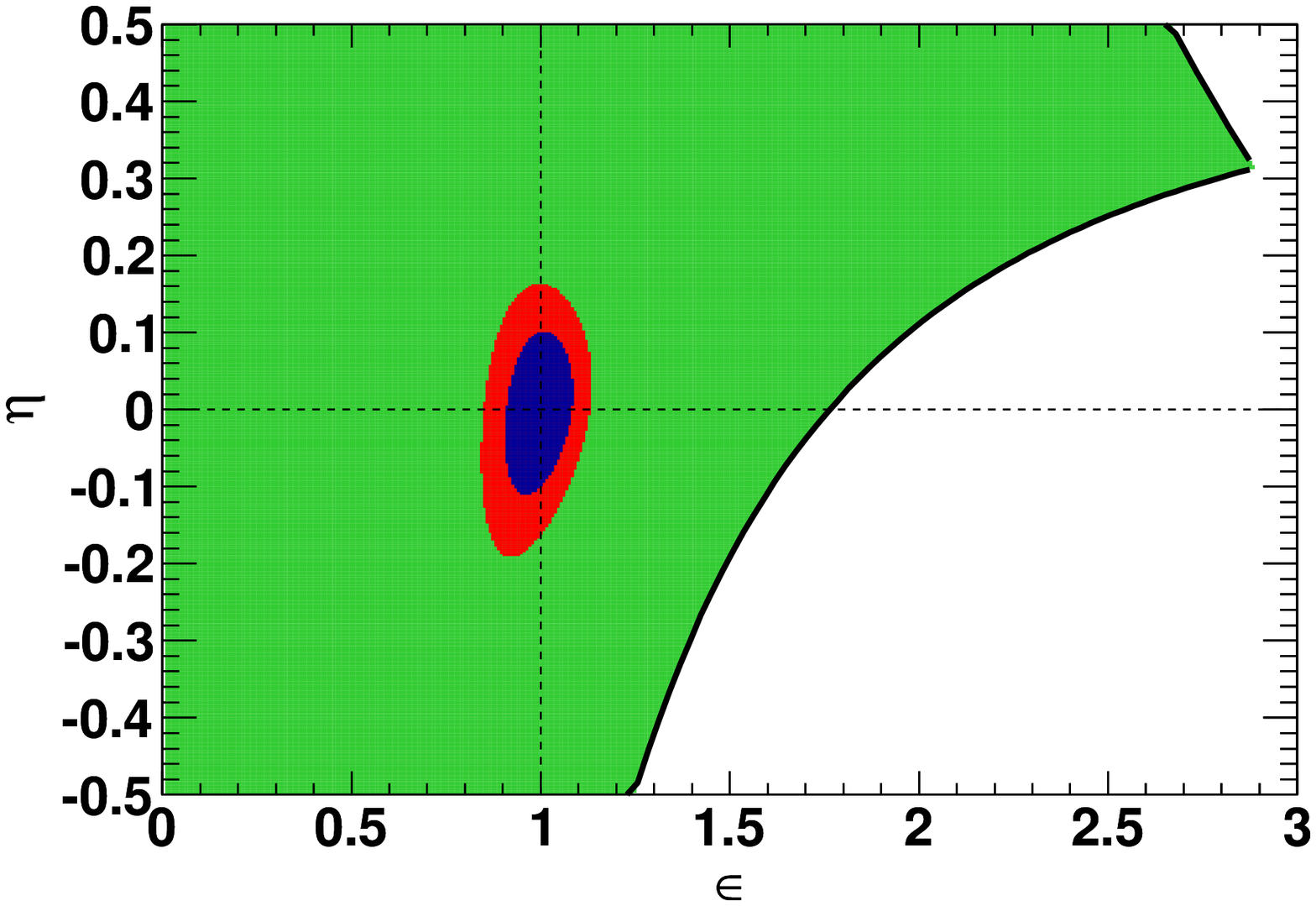}
\caption{\em Expected sensitivities for the ${\rm Br}(B\rightarrow K\nu\bar{\nu})$ (left) and ${\rm Br}(B\rightarrow K^*\nu\bar{\nu})$ (middle)
as a function of the integrated luminosity; and expected constraint on the $(\epsilon,\eta)$ plane for the measurement of the before
mentioned ${\rm Br}$'s at $75{\rm ab^{-1}}$ (right).
}
\label{fig:EpsEta_constrain}
\end{figure}

The SuperB fast simulation has been used to produce signal samples in the before mention detector configurations: BaBar, 
SuperB base-line and SuperB+Fwd-PID. This test showed a $15\%$ to $20\%$ increase in efficiency using the SuperB+Fwd-PID 
configuration with respect to BaBar, depending on the final state, mainly due to the boost reduction. For the time 
being no generic $B\bar{B}$ samples has been produced. To be conservative it has been assumed that the background 
efficiency increases by the same factor as the signal in such a way that the signal to background ratio ($S/B$) stays 
constant. This global increase in efficiency provides a gain on $S/\sqrt{(S+B)}$, which would be interpreted as the 
signal significance for a cut and count analysis. The $S/\sqrt{(S+B)}$ ratio, for both $B\rightarrow K\nu\bar{\nu}$ and 
$B\rightarrow K^*\nu\bar{\nu}$ modes, as a function of the integrated luminosity for the three detector configurations is 
shown in the left and middle plots of figure~\ref{fig:EpsEta_constrain} (BaBar (solid-black), SuperB (dotted-black) and 
SuperB+Fwd-PID (solid-red)). A sensitivity of $15\%$ and $17\%$ are expected for the measurement of the ${\rm Br}(B\rightarrow K\nu\bar{\nu})$ 
and ${\rm Br}(B\rightarrow K^*\nu\bar{\nu})$, respectively, at $75{\rm ab^{-1}}$ for the SuperB+Fwd-PID setup.

The rightmost plot of figure~\ref{fig:EpsEta_constrain} shows the constraint at $68\%$ (blue-region) and $95\%$ (red-region) 
in the $(\epsilon,\eta)$ plane for the expected sensitivities on ${\rm Br}(B\rightarrow K\nu\bar{\nu})$ and 
${\rm Br}(B\rightarrow K^*\nu\bar{\nu})$ at $75{\rm ab^{-1}}$. As can be seen, SuperB promises to significantly reduce the 
NP parameter space.

\section{Summary and outlook}

In summary, it has been investigated the reach of SuperB in the search of the $B\rightarrow K^{(*)}\nu\bar{\nu}$ 
decays with both the hadronic and semi-leptonic techniques. Preliminary results based on the SuperB fast simulation have shown 
an $15$ to $25\%$ increase in the global efficiency with respect to the BaBar. It has also be shown that SuperB will allow 
an unprecedent reduction of the NP parameter space, $(\epsilon,\eta)$ plane, for the expected sensitivities at $75{\rm ab^{-1}}$ 
of data. An angular analysis for the $B\rightarrow K^{(*)}\nu\bar{\nu}$ decay will also be feasible, and the additional observable 
($\langle F_L\rangle$) promises to reduce further the NP parameter space.


\end{document}